\documentclass[journal]{IEEEtran}

\usepackage{amsmath}

\usepackage{amsfonts}  
\usepackage{amssymb}   
\usepackage{algorithmic}
\usepackage{algorithm}
\usepackage{array}
\usepackage{stfloats}
\usepackage{siunitx}
\usepackage{url}
\usepackage{verbatim}
\usepackage{amsmath}
\usepackage[compatibility=false]{caption}
\usepackage{cite}

\usepackage{makecell}
\setcellgapes{1pt} 
\makegapedcells

\usepackage{amsmath}  
\newcommand{\PoCA}{\ensuremath{\mathcal{P}\text{o}CA}}

\usepackage[inkscapelatex=false]{svg}

\usepackage[normalem]{ulem}

\usepackage{xcolor} 
\newcommand{\rev}[1]{\textcolor{black}{#1}} 

\usepackage[hidelinks,colorlinks=true]{hyperref}
\usepackage{threeparttable}

\usepackage{caption}

\usepackage{multirow}
\usepackage{subcaption}

\usepackage{graphicx} 

\newcolumntype{P}[1]{>{\centering\arraybackslash}p{#1}}

\usepackage{cite}
\usepackage{booktabs}

\begin{document}

\title{A Physics-\rev{Inspired} Deep Learning Framework \rev{with Polar Coordinate Attention} for Ptychographic Imaging}

\author{Han Yue, Jun Cheng, \IEEEmembership{Senior Member, IEEE}, Yu-Xuan Ren, Chien-Chun Chen, Grant A. van Riessen, Philip Heng Wai Leong, \IEEEmembership{Senior Member, IEEE} and Steve Feng Shu, \IEEEmembership{Senior Member, IEEE} 
\thanks{Manuscript received April xx, 2024; revised April xx, 2024. 

Han Yue is with the Academy for Engineering \& Technology, Fudan University, Shanghai 200433, China (E-mail: hyue23@m.fudan.edu.cn).

Jun Cheng is with the Institute for Infocomm Research, Agency for Science, Technology and Research (A*STAR), Singapore 138632, Singapore (E-mail: cheng\_jun@i2r.a-star.edu.sg).

Yuxuan Ren is with the Institute for Translational Brain Research, Fudan University, Shanghai 200032, China (E-mail: yxren@fudan.edu.cn).

Chien-Chun Chen is with the Department of Engineering and System Science, National Tsing Hua University, Hsinchu 300044, Taiwan (E-mail: ccchen0627@gmail.com).

Grant A. van Riessen is with the Department of Mathematical and Physical Sciences, School of Computing, Engineering and Mathematical Sciences, La Trobe University, Bundoora, VIC 3086, Australia (E-mail: g.vanriessen@latrobe.edu.au).

Philip H.W. Leong is with the School of Electrical and Computer Engineering, The University of Sydney, Camperdown, NSW 2006, Australia (E-mail: philip.leong@sydney.edu.au).

Steve Feng Shu is with the School of Electrical and Computer Engineering, The University of Sydney, Camperdown, NSW 2006, Australia (E-mail: steve.shu@sydney.edu.au).

Our code will be available at: \href{https://github.com/johncolddd/PPN}{https://github.com/johncolddd/PPN}
 }

}

\markboth{IEEE TRANSACTIONS ON COMPUTATIONAL IMAGING,~Vol.~xx, No.~x, xxxx~2024}%
{Yue \MakeLowercase{\textit{et al.}}:}


\maketitle

\begin{abstract}
Ptychographic imaging confronts inherent challenges in applying deep learning for phase retrieval from diffraction patterns. Conventional neural architectures, both convolutional neural networks and Transformer-based methods, are optimized for natural images with Euclidean spatial neighborhood-based inductive biases that exhibit geometric mismatch with the concentric coherent patterns characteristic of diffraction data in reciprocal space. In this paper, we present \rev{PPN}, a physics-inspired deep learning network with Polar Coordinate Attention (PoCA) for ptychographic imaging, that \rev{aligns neural inductive biases with diffraction physics} through a dual-branch architecture separating local feature extraction from non-local coherence modeling. It consists of a PoCA mechanism that replaces Euclidean spatial priors with physically consistent radial-angular correlations. PPN outperforms \rev{existing end-to-end models, with spectral and spatial analysis} confirming its greater preservation of high-frequency details. \rev{Notably, PPN maintains robust performance compared to iterative methods even at low overlap ratios —  well-suited for high-throughput imaging in real-world acquisition scenarios for samples with consistent structural characteristics.}
\end{abstract}


\section{Introduction}

\IEEEPARstart{C}{oherent} diffraction imaging (CDI) has enabled high-resolution, lens-less imaging across various scientific disciplines by exploiting the principles of wave propagation and interference. In CDI, detectors capture only the far-field intensity distribution of scattered coherent radiation, resulting in the loss of crucial phase information due to the well-known phase problem in crystallography. The objective of phase retrieval algorithms is to reconstruct the complete complex-valued exit wave function from these incomplete Fraunhofer diffraction patterns. Ptychography \cite{hegerlDynamischeTheorieKristallstrukturanalyse1970} , an advanced CDI technique that operates in both real and reciprocal space, addresses this inverse problem by utilizing multiple overlapping diffraction measurements in reciprocal space, effectively extending the Fourier domain sampling and enabling robust phase retrieval through iterative algorithms. This approach offers extended field-of-view imaging with exceptional spatial resolution in real space. Recent breakthroughs have pushed the boundaries of resolution, achieving \SI{0.39}{\angstrom} in transmission electron microscopy \cite{jiangElectronPtychography2D2018} and even 14pm through local-orbital ptychography \cite{yangLocalorbitalPtychographyUltrahighresolution2024}, opening new possibilities in materials characterization \cite{pedersenImprovingOrganicTandem2015}, biological imaging \cite{kasprowiczCharacterisingLiveCell2017}, and semiconductor research\cite{hollerHighresolutionNondestructiveThreedimensional2017}.

\begin{figure}[htbp]
\centering
\includegraphics[width=\linewidth]{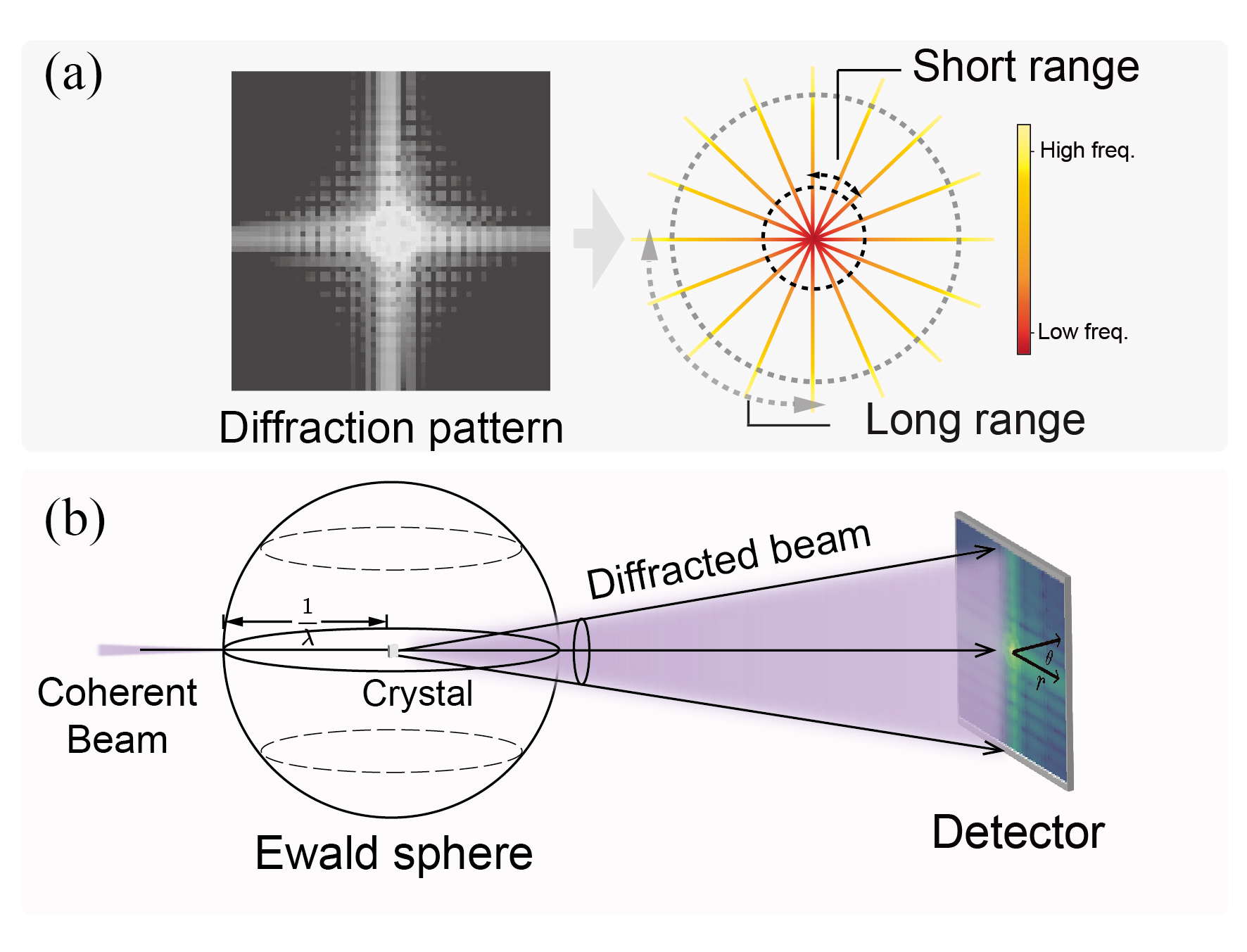}
\caption{Diffraction pattern characteristics and diffraction physics. (a): Diffraction patterns exhibit a radial distribution of information, with varying requirements for capturing low and high-frequency information. (b): 2D diffraction patterns can be viewed as projections of the intersection between the Ewald sphere and the crystal onto the detector plane. Sphere's radius is $1/\lambda$, the wavelength of the incident and diffracted beams. Importantly, the spatial adjacency preference observed in natural images' feature space also holds in the polar coordinate perspective of diffraction patterns.}
\label{fig:imaging_process}
\end{figure}

However, the widespread application of ptychography faces significant challenges, particularly in computational efficiency. As imaging capabilities advance, data volume grows exponentially, overwhelming conventional algorithms. For instance, processing one second of data from a modern synchrotron source (10-megapixel detector, 32-bit depth, 2 kHz, 640 Gb/s) can take up to an hour \cite{babuDeepLearningEdge2023b}. Additionally, achieving high-quality retrievals requires 60-70\% probe position overlap \cite{edoSamplingXrayPtychography2013}, further increasing computational complexity. These factors severely constrain ptychography's viability in real-time and high-throughput applications.

To address these limitations, researchers have redefined ptychography as a data-driven supervised learning task, leveraging deep learning (DL) techniques. Notable examples including PtychoNN \cite{cherukaraAIenabledHighresolutionScanning2020a}, Deep-phase-imaging (DPI) \cite{changDeepLearningElectronDiffractive2023a}, and PtyNet \cite{panEfficientPtychographyReconstruction2023a} have demonstrated improved retrieval efficiency. Unlike conventional approaches that require repeated iterations across the spatial and frequency domains, these methods utilize experimentally acquired diffraction patterns and complex amplitude images reconstructed by traditional algorithms as training data. Through convolutional neural network (CNN) based architectures, they establish a direct mapping from diffraction patterns to complex amplitude retrieval, significantly enhancing computational efficiency. Despite their success, CNN-based methods face challenges in processing multi-frequency information, capturing long-range correlations, and distinguishing high-frequency signals from noise in diffraction patterns, as illustrated in Fig.~\ref{fig:imaging_process}(a). These limitations are due to the sliding window operation of convolution kernels in spatial coordinates and the information flattening in multi-layer convolution during transmission \cite{dengLearningSynthesizeRobust2020, fanPhaseRetrievalBased2024, pelzOntheflyScansXray2014}. Transformer models, while powerful in various computer vision tasks, have limited application in ptychography due to the mismatch between their design principles and the radial nature of diffraction patterns. For instance, Vision Transformers (ViT), are primarily optimized for natural images in real space (e.g., patch extraction operations and local feature preference priors of the Softmax function). However, these priors do not directly map to the radial nature of diffraction data in reciprocal space  shown in Fig.~\ref{fig:imaging_process}(a).

Addressing the inherent limitations of existing methods  requires a fundamental reconsideration of deep learning architectures for ptychographic retrieval, with a focus on optimizing neural networks for representing diffraction patterns in reciprocal space. This insight inspired the development of PPN, a novel framework that bridges frequency-domain diffraction patterns and real-space images using physics-informed deep learning techniques. Drawing inspiration from the Ewald construction in X-ray crystallography, as depicted in Fig.~\ref{fig:imaging_process}(b), treating 2D diffraction patterns as projections of the intersections between the Ewald sphere and the reciprocal lattice onto the detector plane, as shown in the left of Fig.~\ref{fig:imaging_process}. This insight revealed that diffraction patterns are inherently defined in spherical rather than planar geometry. Consequently, we developed Polar Coordinate Attention (PoCA), an attention mechanism leveraging polar coordinates that naturally aligns with diffraction physics. This polar coordinate-based attention mechanism reframes real-space structural priors in the diffraction context, mapping scattering vector magnitude to $r$ and angular information to $\theta$, thus capturing both radial intensity attenuation and angular coherence. Recognizing the multi-scale and multi-frequency nature of diffraction patterns, as shown in the right side of Fig.~\ref{fig:imaging_process}, PPN employs a dual-branch architecture combining a Local Dependencies Branch constructed from standard ViT blocks with a NonLocal Coherence block containing our designed PoCA. This design balances the capture of long-range and local dependencies in diffraction patterns. PPN's architecture intrinsically aligns with ptychography physics, creating a more natural correspondence between model operations and underlying physical processes.

This structural redesign based on physical insights enables our model to more effectively extract and utilize global coherent information and long-range dependencies in diffraction data, thus offering a physics-informed approach to ptychographic retrieval. Our comprehensive evaluation across simulated and real experimental datasets demonstrates PPN's multi-faceted advantages: (1) Superior retrieval quality across all metrics against leading end-to-end baselines, particularly in high-frequency preservation as evidenced through power spectral density curves and cross-sectional intensity profiles; \rev{ (2) Remarkable operational viability - maintaining $<$5\% performance degradation at 30\% overlap ratio while achieving $>$1,000× faster inference than iterative method when tested on samples without feature distribution shifts, crucial for time-sensitive synchrotron experiments; (3) Unprecedented efficiency with 11× fewer parameters than transformer-based counterparts (6.1M vs 68.9M) and Floating Point Operations Per Second (FLOPs) comparable to lightweight CNNs, enabling deployment  in real-world acquisition scenarios. These improvements could significantly enhance the applicability of ptychography in time-sensitive or radiation-sensitive imaging scenarios across various scientific disciplines.}

The primary contributions of this work are summarized as follows:
\begin{itemize}
\item We present PPN, a physics-inspired dual-branch framework specifically designed for ptychographic imaging that addresses the geometric mismatch between Euclidean spatial priors and concentric coherent patterns in reciprocal space.
\item We propose the PoCA mechanism that replaces spatial neighborhood priors with radial-angular correlations, achieving superior high-frequency preservation compared to other end-to-end baselines.
\rev{\item We demonstrate PPN's practical advantages with $>$1000× faster inference than iterative method, $<$5\% performance degradation at 30\% overlap ratio when tested on samples without feature distribution shifts.}
\end{itemize}

\section{Related Works}

\subsection{Deep Learning-Based Ptychographic Imaging}
Deep learning in ptychographic imaging has recently enhanced computational efficiency and retrieval quality, categorized into three strategies:

\subsubsection{Pre-processing}
Integrating deep learning with iterative algorithms improves initial estimates. The physics-informed automatic differentiation ptychography (ADP) framework uses pre-trained autoencoders to map high-dimensional image data to a low-dimensional latent space \cite{seifertNoiserobustLatentVector2024}, while the double deep image prior (DDIP) method reduces the optimization parameter space \cite{duUsingModifiedDouble2021}, both enhancing convergence rates and noise robustness. \rev{However, these methods still require traditional iterative algorithms.}

\subsubsection{Post-processing}
Neural networks refine reconstructions from traditional algorithms, including enhancing a single iteration of the iterative algorithm to improve spatial resolution and reduce artifacts \cite{kasprowiczCharacterisingLiveCell2017} \cite{ganPtychoDVVisionTransformerBased2024}. \rev{While these methods significantly improve the quality of reconstructed images, they still rely on initial reconstructions and cannot provide real-time or fully automated solutions.}

\subsubsection{End-to-End \rev{with CNNs}}
End-to-end methods directly map diffraction patterns to complex object functions, bypassing iterative processes. Examples include PtychoNN with a modified U-Net and two-branch decoder for amplitude and phase \cite{cherukaraAIenabledHighresolutionScanning2020a}, PtyNet with group convolution and Leaky ReLU for efficiency \cite{panEfficientPtychographyReconstruction2023a}, and DPI using a traditional U-Net with skip connections \cite{changDeepLearningElectronDiffractive2023a}. 
Whereas CNNs excel at local feature extraction for natural images\cite{simonyanVeryDeepConvolutional2015, krizhevskyImagenetClassificationDeep2012}, their inductive biases prove suboptimal for diffraction patterns requiring global phase coherence. The quadratic phase factors in Fresnel propagation create position-dependent correlations that span the entire detector plane, yet the local receptive fields of CNNs (typically 3×3) cannot span the full radial extent of diffraction rings, causing these physically critical phase relationships between distant pixels to be irrecoverably lost during feature encoding. Furthermore, progressive downsampling in hierarchical architectures systematically discards high-frequency information during feature abstraction. These inherent limitations of CNNs motivate the exploration of attention mechanisms for better long-range dependency modeling.

\subsection{\rev{Transformer-based Methods}}
\begin{figure*}[htbp]
\centering
\includegraphics[width=\linewidth]{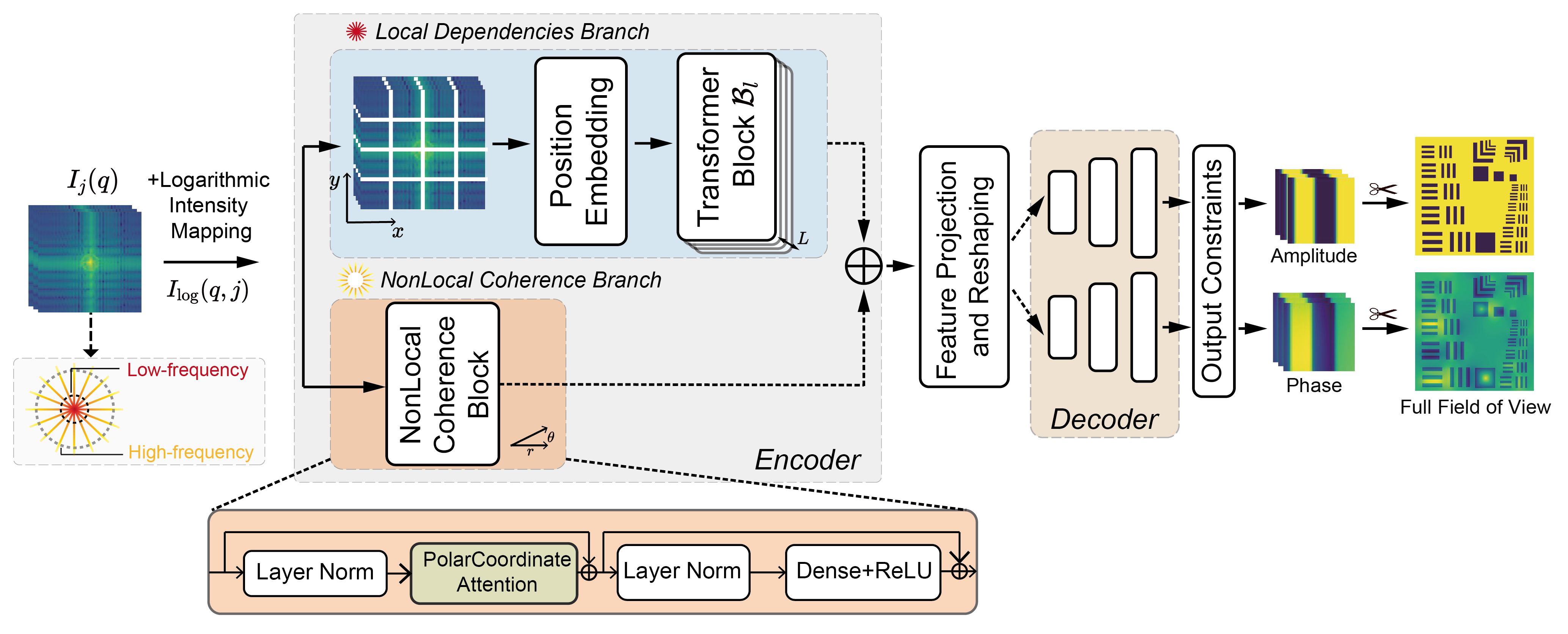}
\caption{The proposed PPN for ptychographic imaging. It features dual branches: a Local Dependencies Branch with standard ViT blocks, and a NonLocal Coherence Branch with  Polar Coordinate Attention mechanism. The model processes logarithmically mapped diffraction patterns, combining features from both branches before decoding into individual amplitude and phase reconstructions \rev{at each position, which are then separately stitched to generate full field of view images for both amplitude and phase.}}
\label{fig:moxing}
\end{figure*}

Vision Transformers (ViTs) \cite{dosovitskiyImageWorth16x162021} emerged as a promising solution to the limitations of CNNs, demonstrating exceptional capability in capturing long-range dependencies across various domains \cite{liMvitv2ImprovedMultiscale2022, hanSurveyVisionTransformer2022, wuMedsegdiffv2DiffusionbasedMedical2024,nguyenClimatelearnBenchmarkingMachine2024, changBidirectionalGenerationStructure2024}. Recent adaptations in ptychography, such as PtychoFormer\cite{nakahataPtychoFormerTransformerbasedModel2024} employs a hybrid architecture combining ViT and CNN components, specifically a SegFormer variant adapted for ptychography. While it surpasses traditional CNN methods in reconstruction accuracy, this comes with significant increases in parameter count and training costs. \rev{PtychoDV \cite{ganPtychoDVVisionTransformerBased2024} adopts a compromise strategy that utilizes ViTs' reconstruction results as initial guesses for iterative methods, thereby enhancing the final accuracy of conventional iterative approaches. This indirectly reveals that pure end-to-end deep learning methods still face performance bottlenecks in diffraction reconstruction tasks - the core issue we aim to address in this paper.}

While Transformers demonstrate superior capability in capturing global contexts, their direct application to diffraction patterns neglects crucial physical priors inherent in reciprocal space representations.
Diffraction patterns, governed by wave optics, approximate the Fourier transform of an object's transmission function in far-field conditions \cite{parrentFraunhoferFarField1964}, resulting in reciprocal space features like sparse representations and concentric rings \cite{buchwaldKirchhoffTheoryOptical2016}. Accurate image retrieval and phase retrieval require effectively capturing both sparsity and long-range dependencies within these patterns. Although Vision Transformers achieve global modeling through multi-head self-attention mechanisms (MHSA), their Euclidean spatial neighborhood-based attention weight calculation exhibits geometric mismatch with the concentric coherent patterns in reciprocal space characteristic of diffraction patterns. \rev{Our focus is to resolve the critical bottleneck of effective information extraction from diffraction patterns and improved reconstruction of high-frequency spatial information, rather than surpassing or replacing the ultimate resolution achieved by iterative algorithms.}

\section{Methodology}

\subsection{Problem Formulation}

In ptychography, we aim to reconstruct a complex-valued object function $O(\mathbf{r})$, where $\mathbf{r}$ is the position vector in real space. This retrieval is based on a set of diffraction patterns measured at spatial frequencies $\mathbf{q}$ in reciprocal space. The process involves a probe function $P(\mathbf{r})$, which interacts with the object at various scanning positions $\{\mathbf{r}_j\}_{j=1}^J$. The ptychography problem can be formulated as:

\begin{equation}
\mathcal{F}\{\psi_j(\mathbf{r})\} = \sqrt{I_j(\mathbf{q})} \cdot \exp(i\phi_j(\mathbf{q}))
\label{eq:ptychography_problem}
\end{equation}
where $\psi_j(\mathbf{r}) = P(\mathbf{r} - \mathbf{r}_j) \cdot O(\mathbf{r})$ is the exit wave function, $\mathcal{F}$ denotes the Fourier transform, $I_j(\mathbf{q})$ is the measured diffraction intensity, and $\phi_j(\mathbf{q})$ is the phase of the diffraction pattern. To account for real-world factors, we model the diffraction intensity as:

\rev{\begin{equation}
\begin{split}
I_j(\mathbf{q}) = &\text{Poisson}\Big( \eta \Big| \mathcal{F} \Big\{ \mu P(\mathbf{r} - \mathbf{r}_j - \delta\mathbf{r}_j) O(\mathbf{r}) \\
&+ (1 - \mu) P(\mathbf{r} - \mathbf{r}_j - \delta\mathbf{r}_j)\cdot \mathbb{E}_{\mathbf{r}}[O(\mathbf{r})] \Big\} \Big|^2 \Big) \\
&+ \mathcal{N}(0, \sigma^2)
\end{split}
\label{eq:2}
\end{equation}}

where $\mu$ is the coherence parameter, \rev{$\eta \in [0,1]$} is the detector efficiency representing the quantum efficiency (probability of detecting each photoelectron), $\text{Poisson}(\lambda)$ denotes the Poisson distribution with rate parameter $\lambda = \eta \cdot \left| \mathcal{F} \left\{ \dots \right\} \right|^2$, where $\left| \mathcal{F} \left\{ \dots \right\} \right|^2$ is the theoretical diffraction intensity (in photons/pixel) and their product forms the expected value of the Poisson process, and $\delta\mathbf{r}_j$ accounts for positional jitter.

\rev{Based on the physical model of the actual imaging process in Eq.~\eqref{eq:2}, we formulate the objective function for ptychographic retrieval in Eq.~\eqref{eq:objective_functional} to effectively recover the object and probe functions:}
\begin{equation}
\begin{split}
\mathcal{J}(O, P) = &\frac{1}{J}\sum_{j=1}^J \left\|I_j - |\mathcal{F}\{P(\mathbf{r} - \mathbf{r}_j) \cdot O(\mathbf{r})\}|^2\right\|_2^2 \\
&+ \lambda_1 \Omega_1(O) + \lambda_2 \Omega_2(P)
\end{split}
\label{eq:objective_functional}
\end{equation}
where $\left\| \mathbf{x} \right\|_2^2 = \sum_i x_i^2$ is the squared Euclidean norm, $\Omega_1(O) = \left\| \nabla O \right\|_{1}$ enforces sparsity in the object gradient, $\Omega_2(P) = \|P - P_0\|_F^2$ constrains the probe function to maintain proximity to to an initial estimate $P_0$, and $\lambda_1$ and $\lambda_2$ are regularization weights.

\subsection{PPN Architecture}
The proposed architecture employs a bifurcated structure to map diffraction patterns to a reconstructed complex-valued object function. The input undergoes a logarithmic transformation $I_{\text{log}}(\mathbf{q}, j) = \log(1 + I_j(\mathbf{q}))$ as a preprocessing step to address the large disparity between high and low frequency values in the diffraction patterns. 

\subsubsection{Local Dependencies Branch}
The Local Dependencies Branch utilizes a standard Vision Transformer (ViT) to analyze diffraction patterns, capturing local features and structural relationships within each pattern. Operating on input tensors $\mathbf{X} \in \mathbb{R}^{B \times H \times W \times C}$ (where $B$ is batch size, $H \times W$ is spatial resolution, and $C$ is channel depth), this branch processes data through several key components:

Step 1. Patch Extraction divides the input image into non-overlapping patches of size $P \times P$, resulting in \rev{ $\mathbf{X}_{\text{patches}} \in \mathbb{R}^{B \times N_p \times (P^2 \cdot C)}$ where $N_p = \frac{H}{P} \times \frac{W}{P}$ represents the total number of patches.}

Step 2. Linear Projection maps each patch to a $D$-dimensional embedding space using 
   $\mathbf{X}_{\text{embed}} = \mathbf{X}_{\text{patches}} \mathbf{W}_E \in \mathbb{R}^{B \times N_p \times D}$ 
   where $\mathbf{W}_E \in \mathbb{R}^{(P^2C) \times D}$ is the projection matrix.

Step 3. Positional Encoding adds spatial information via 
$\mathbf{X}_{\text{pos}} = \mathbf{X}_{\text{embed}} + \mathbf{E}_{\text{pos}}$, 
where the positional encoding matrix $\mathbf{E}_{\text{pos}} \in \mathbb{R}^{N_p \times D}$ 
is broadcasted across the batch dimension to match $\mathbf{X}_{\text{embed}} \in \mathbb{R}^{B \times N_p \times D}$.

Step 4. Transformer Blocks with Pre-LN structure process the sequence through $L$ layers according to $\mathbf{X}'^{(l)} = \mathcal{MSA}(\mathcal{LN}(\mathbf{X}^{(l-1)})) + \mathbf{X}^{(l-1)}$ and $\mathbf{X}^{(l)} = \mathcal{FFN}(\mathcal{LN}(\mathbf{X}'^{(l)})) + \mathbf{X}'^{(l)}$. The Multi-Head Self-Attention (MSA) mechanism computes $\mathcal{MSA}(\mathbf{X}) = \text{Concat}(\text{head}_1,...,\text{head}_h)\mathbf{W}^O$ where each attention head captures different relationship patterns in the data, with projection matrices $\mathbf{W}^Q_i, \mathbf{W}^K_i \in \mathbb{R}^{D \times d_k}$ and $\mathbf{W}^V_i \in \mathbb{R}^{D \times d_v}$ (typically $d_k = d_v = D/h$). 

Step 5. Spatial Restoration reshapes the output to $\mathcal{H}_{\text{Local}} \in \mathbb{R}^{B \times (H/P) \times (W/P) \times D}$.

\rev{The complete dimension flow is: $\mathbb{R}^{B \times H \times W \times C} 
   \xrightarrow{\text{(1)}} \mathbb{R}^{B \times N_p \times (P^2C)} 
   \xrightarrow{\text{(2)}} \mathbb{R}^{B \times N_p \times D} 
   \xrightarrow{\text{(3)}} \mathbb{R}^{B \times N_p \times D} \xrightarrow{\text{(4)}} \mathbb{R}^{B \times N_p \times D} \xrightarrow{\text{(5)}} \mathbb{R}^{B \times (H/P) \times (W/P) \times D}$. }

\subsubsection{Non-Local Coherence Branch}
The Non-Local Coherence Branch employs an architectural framework similar to the Local Dependencies Branch, sharing identical transformer block components while introducing our novel Polar Coordinate Attention (PoCA) mechanism. This specialized attention mechanism processes input at pixel-level granularity ($H \times W$ pixels) without patch extraction, enabling the capture of long-range dependencies across diffraction patterns. The PoCA mechanism systematically encodes physical constraints inherent to diffraction physics, including the $I \propto r^{-4}$ intensity distribution through radial decay weighting, preserves crystallographic symmetry via angular continuity, and adapts to illumination shifts with dynamically learned centering parameters $\alpha_x$ and $\alpha_y$.

After logarithmic transformation, the single-channel diffraction pattern $\mathbb{R}^{B \times H \times W \times 1}$ is projected to a multi-channel tensor $\mathbb{R}^{B \times H \times W \times C}$ via a learnable $1 \times 1$ convolutional layer, where $C$ denotes the embedding dimension. For an input tensor $\mathbf{X} \in \mathbb{R}^{B\times H\times W\times C}$ (where $B$ represents batch size, $H \times W$ spatial resolution), this branch processes data through several key components:

Step 1. Tensor Flattening transforms the input into a sequence representation via $\mathbf{X}_f = \text{Reshape}(\mathbf{X}, (B, H\times W, C)) \in \mathbb{R}^{B \times N \times C}$, where $N=H\times W$ represents the total number of pixels.

Step 2. Query-Key-Value Projection maps the flattened tensor into three separate embedding spaces using:
\begin{equation}
\begin{aligned}
\mathcal{Q}^i &= \mathbf{X}_f \mathbf{W}_Q^i \quad &\mathbf{W}_Q^i &\in \mathbb{R}^{C \times d_k} \\
\mathcal{K}^i &= \mathbf{X}_f \mathbf{W}_K^i \quad &\mathbf{W}_K^i &\in \mathbb{R}^{C \times d_k} \\
\mathcal{V}^i &= \mathbf{X}_f \mathbf{W}_V^i \quad &\mathbf{W}_V^i &\in \mathbb{R}^{C \times d_v}
\end{aligned}
\end{equation}
where projection matrices $\mathbf{W}_Q^i, \mathbf{W}_K^i \in \mathbb{R}^{C \times d_k}$ and $\mathbf{W}_V^i \in \mathbb{R}^{C \times d_v}$ define the embedding transformations for attention head $i$. The resulting tensors are $\mathcal{Q}^i, \mathcal{K}^i \in \mathbb{R}^{B \times N \times d_k}$ and $\mathcal{V}^i \in \mathbb{R}^{B \times N \times d_v}$, where the matrix multiplication is performed along the feature dimension $C$. Polar coordinate parameterization defines the physical diffraction geometry using:

\begin{equation}
c_x = \frac{W}{2} + \alpha_x \frac{W}{2}, \quad
c_y = \frac{H}{2} + \alpha_y \frac{H}{2}
\end{equation}
 \rev{
\begin{equation}
r_m = \frac{\log(1 + \|\mathbf{p}_m - \mathbf{c}\|)}{\log(1 + r_{\max})}, \quad
\theta_m = \arctan2\left( y_m - c_y, x_m - c_x \right)
\end{equation}}
 \rev{
\begin{equation}
\mathbf{\Phi}_r^{mn} = \frac{1}{1 + |r_m - r_n|}, \quad 
\mathbf{\Phi}_\theta^{mn} = \cos(\theta_m - \theta_n)
\end{equation}}
  
where $r_{\max} = \sqrt{(W/2)^2 + (H/2)^2}$ ensures normalized radial coordinates, \rev{$\mathbf{p}_m=(x_m,y_m)$ denotes original pixel coordinates, $\mathbf{c}=(c_x,c_y)$ represents the learned diffraction center, and $\alpha_x, \alpha_y \in [-0.5, 0.5]$ are learnable parameters initialized at 0.} The implementation employs logarithmic scaling to compress the dynamic range of radial distances, with angle normalization $\theta_m \in [0,2\pi)$ ensuring continuity.

Step 3. Polar Coordinate Modulation incorporates physical constraints by modulating the standard dot-product attention with radial and angular weighting matrices:

\begin{equation}
\begin{split}
\mathcal{A}_{\text{polar}}^i &= \underbrace{\frac{\mathcal{Q}^i(\mathcal{K}^i)^\top}{\sqrt{d_k}}}_{\text{base attention}} 
                               \odot \underbrace{\mathbf{\Phi}_r}_{\text{radial decay}} 
                               \odot \underbrace{\mathbf{\Phi}_\theta}_{\text{angular continuity}}
\end{split}
\end{equation}

where $\mathbf{\Phi}_r$, $\mathbf{\Phi}_\theta \in \mathbb{R}^{1 \times N \times N} \quad \text{(broadcastable to } \mathbb{R}^{B \times N \times N}\text{)}$,  $\odot$ \rev{ represents element-wise multiplication with broadcasting, imposing both radial decay and angular continuity constraints.}

Step 4. Attention Application computes each attention head through $\text{Head}_i = \text{softmax}\left( \mathcal{A}_{\text{polar}}^i \right) \mathcal{V}^i$, where the softmax normalization ensures proper probability distribution across the attention weights, and $\mathcal{A}_{\text{polar}}^i \in \mathbb{R}^{B \times N \times N}$ is applied to $\mathcal{V}^i \in \mathbb{R}^{B \times N \times d_v}$.

Step 5. Multi-head Integration combines information from all h attention heads via \rev{ $\mathcal{G}(\mathbf{X}_f) = \text{Concat}(\text{Head}_1,...,\text{Head}_h) \mathbf{W}^O$, with $\mathbf{W}^O \in \mathbb{R}^{h \cdot d_v \times C'}$ serving as the output projection matrix and $\text{Head}_i \in \mathbb{R}^{B \times N \times d_v}$.}

Step 6. Spatial Restoration reshapes the processed output to match the original spatial dimensions through $\PoCA{} (\mathbf{X}) = \text{Reshape}\left( \mathcal{G}(\mathbf{X}_f), (B,H,W,C') \right) \in \mathbb{R}^{B\times H\times W\times C'}$.

\rev{The complete dimension flow through this branch is:
$\mathbb{R}^{B\times H\times W\times C} \xrightarrow{(1)} \mathbb{R}^{B\times N\times C} \xrightarrow{(2)} \{\mathbb{R}^{B\times N\times d_k}$ $(\mathcal{Q}^i, \mathcal{K}^i), \mathbb{R}^{B\times N\times d_v} (\mathcal{V}^i)\} \xrightarrow{(3)} \mathbb{R}^{B\times N\times N} \xrightarrow{\text{softmax}} \mathbb{R}^{B\times N\times N} \xrightarrow{(4)} \mathbb{R}^{B\times N\times d_v} \xrightarrow{(5)} \mathbb{R}^{B\times N\times h \cdot d_v} \xrightarrow{\mathbf{W}^O} \mathbb{R}^{B\times N\times C'} \xrightarrow{(6)} \mathbb{R}^{B\times H\times W\times C'}$}

\rev{By decomposing attention into radial and angular components, PoCA effectively mimics the Ewald sphere's intersection with the reciprocal lattice (Fig.~\ref{fig:imaging_process}(b)), where $r_{mn}$ correlates with the reciprocal space resolution of scattering vectors, and $\theta_{mn}$ encodes Bragg angle relationships through angular coherence constraints. }

\subsubsection{Feature Fusion and Decoding}
The feature fusion process addresses the resolution mismatch between the two branches. The Local Dependencies Branch output $\mathcal{H}_{\text{Local}} \in \mathbb{R}^{B \times (H/P) \times (W/P) \times D}$ has lower spatial resolution than the Non-Local Branch output $\mathcal{H}_{\text{NonLocal}} \in \mathbb{R}^{B \times H \times W \times C'}$. To align these representations, we employ bilinear upsampling on the Local Branch features:

\begin{equation}
\mathcal{H}_{\text{Local}}^{\text{upsampled}} = \text{Upsample}_{\text{bilinear}}(\mathcal{H}_{\text{Local}}) \in \mathbb{R}^{B \times H \times W \times D}
\end{equation}

The aligned features are then combined through concatenation followed by a 1×1 convolution:
\begin{equation}
\mathcal{H}_{\text{fused}} = \text{Conv}_{1\times1}(\text{Concat}[\mathcal{H}_{\text{Local}}^{\text{upsampled}}, \mathcal{H}_{\text{NonLocal}}])
\end{equation}
   
The decoder $\mathcal{D}$ consists of three transposed convolution blocks with progressive upsampling:
\begin{equation}
\begin{split}
\mathcal{D}_i &= \text{BatchNorm}(\text{ReLU}(\text{ConvTranspose2D}_{k,s})) \\
\mathcal{D} &= \mathcal{D}_3 \circ \mathcal{D}_2 \circ \mathcal{D}_1
\end{split}
\end{equation}
where $k$ is kernel size and $s$ is stride. The decoder branches into parallel paths $\mathcal{D}_{\text{amplitude}}$ and $\mathcal{D}_{\text{phase}}$ to reconstruct the object's amplitude and phase components.

\subsubsection{Training Objective}

The loss function $\mathcal{L}: \Theta \to \mathbb{R}^+$ is defined as:

\begin{equation}
\mathcal{L}(\theta) = \frac{1}{N} \sum_{n=1}^N \left[ \left\| A_{\text{true}}^{(n)} - A_{\text{pred}}^{(n)} \right\|_2^2 + \left\| \phi_{\text{true}}^{(n)} - \phi_{\text{pred}}^{(n)} \right\|_2^2 \right]
\label{eq:lossfunction}
\end{equation}

where $N$ is the number of training samples, $A_{\text{true}}^{(n)}$ and $\phi_{\text{true}}^{(n)}$ are the true amplitude and phase for the $n$-th sample (both amplitude and phase values are normalized to the range [0,1]), respectively, and $A_{\text{pred}}^{(n)} = A_{\text{final}}^{(n)}$ and $\phi_{\text{pred}}^{(n)} = \boldsymbol{\phi}_{\text{final}}^{(n)}$ are the corresponding model predictions. \rev{The two error terms naturally maintain comparable scales without requiring additional weighting factors.}The optimization problem is formulated as:

\begin{equation}
\theta^* = \arg\min_{\theta \in \Theta} \mathcal{L}(\theta)
\label{eq:optimizationproblem}
\end{equation}
where $\theta^*$ represents the optimal model parameters that minimize the loss function $\mathcal{L}(\theta)$ over the training dataset.

\section{Experiments and Results}
\subsection{Experiment Setup}

\begin{figure*}[htbp]
\centering
\includegraphics[width=1.05\linewidth]{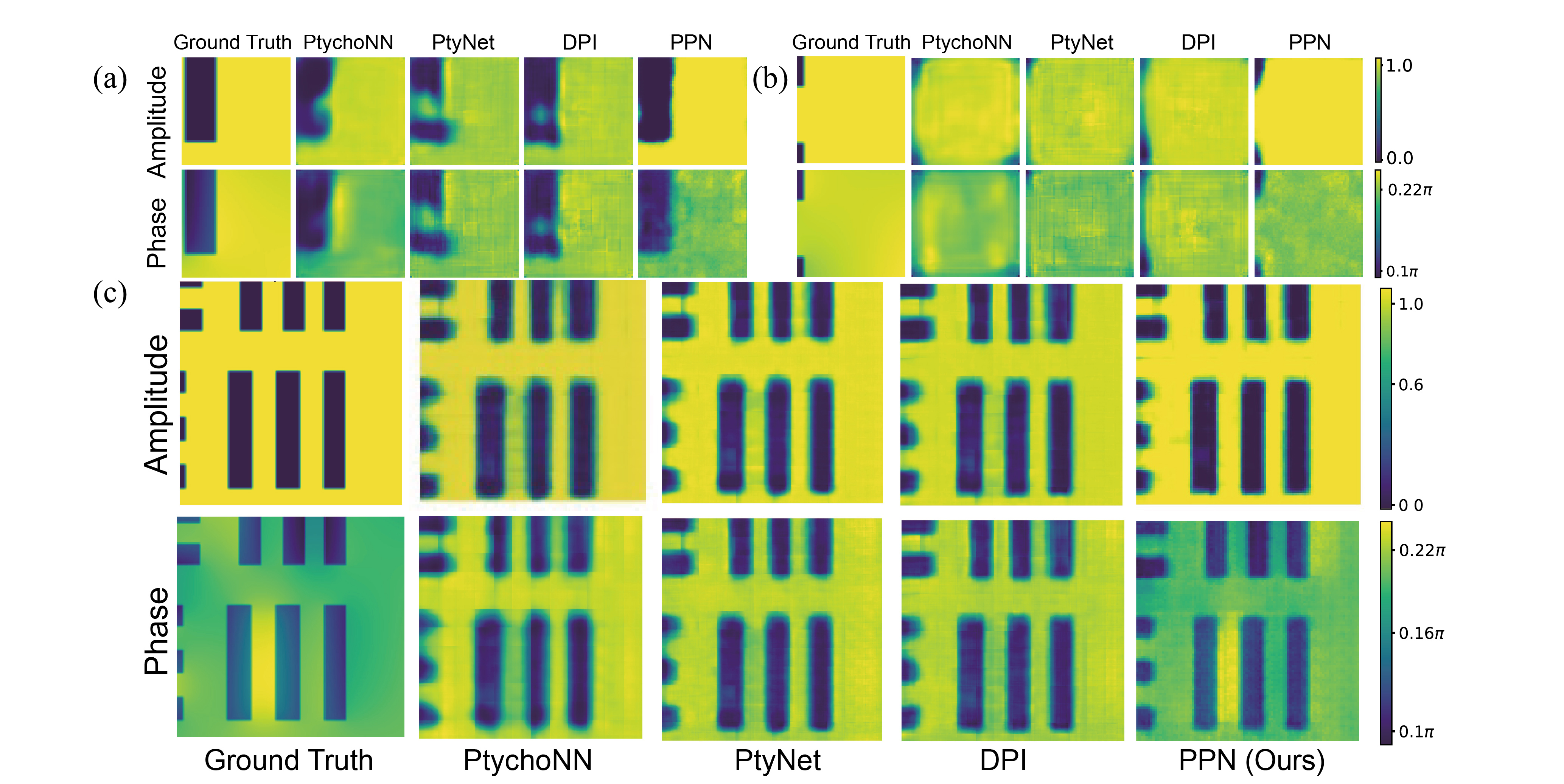}
\caption{Performance comparison of single-shot experiment results and full-stitched scene retrieval using simulated data. (a) and (b) show amplitude and phase reconstructions at two representative scan positions. (c) displays the full-field amplitude and phase images \rev{stitched together from individual reconstructions at all scan positions} across different models.
}
\label{fig:simu}
\end{figure*}

\subsubsection{Implementation Details}
We implemented both our DL-based method and the traditional iterative algorithm (like the extended ptychographic iterative engine (ePIE) algorithm\cite{maidenImprovedPtychographicalPhase2009b}) in Python with CUDA acceleration on the same hardware platform. \rev{The iterative algorithm was implemented with CUDA support, initialized with known probe measurements and processing 32$\times$32 diffraction patterns, with phase unwrapping applied to retrieve the continuous phase. The ePIE implementation typically required 100-200 iterations to converge and 75-80\% spatial overlap, with minimal gains beyond. } Both implementations were trained and evaluated on a \rev{NVIDIA\textsuperscript{\textregistered} Tesla T4 GPU with Intel\textsuperscript{\textregistered} Xeon\textsuperscript{\textregistered} Platinum 8480CL CPU @ 2.00GHz running Linux 6.1.85+.} For the DL-based method, we used TensorFlow with an MSE loss function and the Adam optimizer. We set the initial learning rate to 1.0 and implemented a ReduceLROnPlateau callback (factor 0.5, patience 2, $\min_{\operatorname{lr}}$ 0.0001) to dynamically adjust the learning rate based on validation loss. We used a batch size of 32 with 5\% validation split, while a custom callback monitored system memory usage, computation time, and convergence behavior.

\subsubsection{Datasets}
We utilize two types of datasets in our experiments: \textbf{Simulated Dataset}: The dataset consists of a 512×512 LTEM image\cite{zhouDifferentialProgrammingEnabled2021}. \rev{The pattern is based on a slightly modified version of the 1951 USAF resolution test chart}. Using a probe size of 32×32 pixels with a 75\% overlap rate and 3-pixel jitter, we generated a total of 3721 patches arranged in a 61×61 grid, each patch being 32×32 pixels. Gaussian noise simulates realism, with added Poisson and Gaussian noise for readout and photon effects. Amplitude and phase undergo minmax normalization. \textbf{Real Experimental Data}: Experimental data from Argonne National Laboratory \cite{cherukaraAIenabledHighresolutionScanning2020a} comprises 16,100 triplets of diffraction data, amplitude, and phase images. 161 × 161 point scan at 30 nm increments from X-ray nanoprobe beamline 26-ID. PIE algorithm generates ground truth. For both datasets, we split the data into 80\% for training and 20\% for testing. The validation set is created using 5\% of the training data. Results on simulated data are presented in Sections IV.B, D, E, F, G and Section V.B, while Section IV.C validates the model using real experimental synchrotron data. 

\subsubsection{Evaluation Metrics}
The performance of the models is evaluated using MSE, Peak Signal-to-Noise Ratio (PSNR), and SSIM, with the ground truth values of the materials basis images as references.

\begin{table*}[htbp]
\caption{Quantitative comparison of full-stitched scenes under noise-free and noisy conditions on simulated data. Results are presented as mean $\pm$ standard deviation from ten independent experiments.}
\label{tab:combined_performance}
\centering
\resizebox{\textwidth}{!}{%
\begin{tabular}{llcccccc}
\toprule
\multirow{2}{*}{Condition} & \multirow{2}{*}{Method} & \multicolumn{3}{c}{Amplitude} & \multicolumn{3}{c}{Phase} \\
\cmidrule(r){3-5} \cmidrule(l){6-8}
& & MSE ($\times10^{-2}$) $\downarrow$ & PSNR (dB) $\uparrow$ & SSIM (\%) $\uparrow$ & MSE ($\times10^{-2}$) $\downarrow$ & PSNR (dB) $\uparrow$ & SSIM (\%) $\uparrow$ \\
\midrule
\multirow{4}{*}{Noise-free} 
& PtychoNN\cite{cherukaraAIenabledHighresolutionScanning2020a} & 4.32 $\pm$ 0.02 & 13.52 $\pm$ 0.17 & 83.50 $\pm$ 0.80 & 1.22 $\pm$ 0.01 & 19.09 $\pm$ 0.19 & 67.10 $\pm$ 0.70 \\
& PtyNet\cite{panEfficientPtychographyReconstruction2023a} & 4.46 $\pm$ 0.03 & 13.56 $\pm$ 0.15 & 83.10 $\pm$ 0.60 & 1.28 $\pm$ 0.01 & 19.10 $\pm$ 0.26 & 67.80 $\pm$ 1.30 \\
& DPI\cite{changDeepLearningElectronDiffractive2023a} & 4.29 $\pm$ 0.03 & 13.68 $\pm$ 0.12 & 83.40 $\pm$ 0.50 & 1.26 $\pm$ 0.01 & 18.91 $\pm$ 0.24 & 67.70 $\pm$ 1.10 \\
& PPN (Ours) & \textbf{3.63 $\pm$ 0.04} & \textbf{14.42 $\pm$ 0.08} & \textbf{87.00 $\pm$ 0.30} & \textbf{1.14 $\pm$ 0.01} & \textbf{19.41 $\pm$ 0.05} & \textbf{70.60 $\pm$ 0.70} \\
\midrule
\multirow{4}{*}{Noisy} 
& PtychoNN & 6.66 $\pm$ 0.03 & 11.76 $\pm$ 0.12 & 74.00 $\pm$ 0.89 & 1.47 $\pm$ 0.04 & 18.33 $\pm$ 0.17 & 58.90 $\pm$ 1.00 \\
& PtyNet & 6.50 $\pm$ 0.02 & 11.91 $\pm$ 0.13 & 74.80 $\pm$ 0.66 & 1.43 $\pm$ 0.07 & 18.46 $\pm$ 0.18 & 62.10 $\pm$ 1.50 \\
& DPI & 6.46 $\pm$ 0.02 & 11.76 $\pm$ 0.10 & 74.20 $\pm$ 0.65 & 1.48 $\pm$ 0.06 & 18.31 $\pm$ 0.20 & 62.00 $\pm$ 1.20 \\
& PPN (Ours) & \textbf{5.91 $\pm$ 0.03} & \textbf{12.34 $\pm$ 0.11} & \textbf{78.80 $\pm$ 0.59} & \textbf{1.31 $\pm$ 0.06} & \textbf{18.76 $\pm$ 0.16} & \textbf{68.30 $\pm$ 1.10} \\
\bottomrule
\end{tabular}
}
\caption*{\footnotesize{Note: For noisy conditions, noise is simulated using Gaussian and Poisson distributions to model readout noise and photon noise, respectively. Bold values indicate the best performance for each metric. PSNR in dB, and SSIM in percentage.  $\uparrow$: higher is better, $\downarrow$: lower is better.}}
\end{table*}

\subsection{Performance Comparison and Analysis}
\label{sec:performance}
We evaluated PPN against three mainstreamed CNN-based methods  (PtychoNN, PtyNet, and DPI). Our analysis covered single-shot retrieval and full-stitched field retrieval under both ideal and noisy conditions.

\subsubsection{Single Scan Point Retrieval Analysis}
PPN reconstructed retrievals with improved edge definition and clarity compared to those from the other methods \rev{in single scan point recovery, where each position represents an individual amplitude/phase reconstruction from a corresponding diffraction pattern.} This improvement is particularly evident in the vertical edge on the left side of Fig. \ref{fig:simu}(a). The reconstructed images also showed contrast levels closer to the ground truth, with more natural transitions between dark and bright areas. By contrast, the CNN-based models tended to smooth these sharp features and introducing artifacts and distortions accuracy.

\begin{figure}[htbp]
\centering

\includegraphics[width=0.9\linewidth]{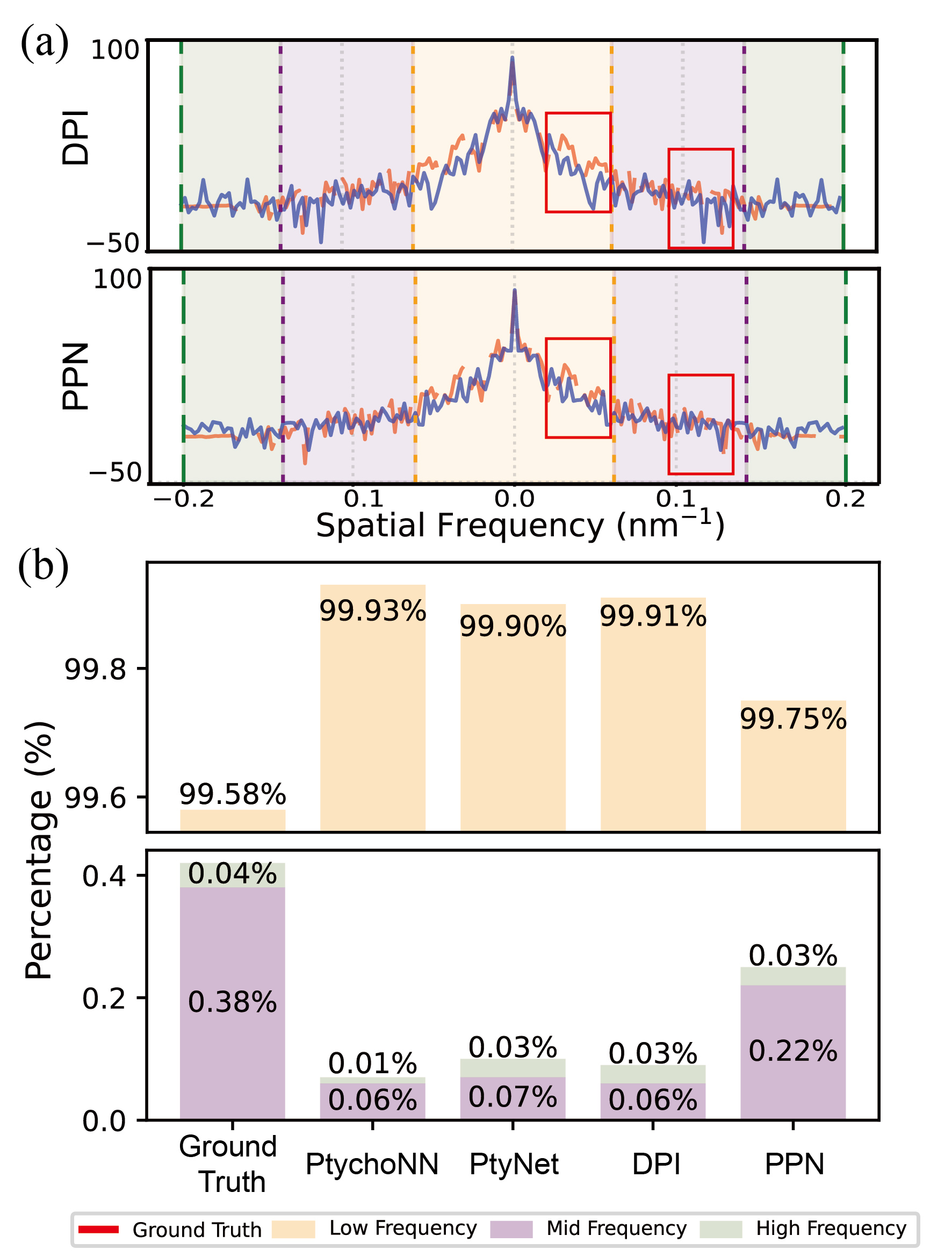}
\caption{Frequency analysis comparison of full-scene ptychographic retrievals on simulated data. (a) 1D diagonal cross-sections of average 2D PSD from stitched simulations. Red curve: ground truth; Blue curve: models. The red box represents obvious abnormal model retrieval. (b) Quantitative breakdown of energy distribution across low, mid, and high frequency bands. Frequency ranges are defined based on the radial distance from the PSD center, with boundaries at 1/3 and 2/3 of the maximum frequency. 
}

\label{fig:psd}
\end{figure}

\subsubsection{Full-stitched Field Retrieval Analysis}\label{subsec:full_stitched_field_retrieval}
Table \ref{tab:combined_performance} presents full-field stitching results under ideal (noise-free) and realistic (noisy) conditions, demonstrating PPN's superior performance across all metrics. In ideal conditions, PPN achieved 16.0\% lower MSE\textsubscript{amp} and 6.7\% higher PSNR\textsubscript{amp} compared to PtychoNN, with similar improvements under noisy conditions. Fig. \ref{fig:simu}(c) visually confirms PPN's superior global consistency and minimal boundary artifacts, particularly in reconstructing vertical column ends and background textures. The dual-branch structure enables our model to effectively capture both low-frequency and mid-to-high frequency information of diffraction patterns. This capability stems from the model's ability to capture the intrinsic geometric structure of the data, analogous to continuous mapping on high-dimensional manifolds.

Repeated measures ANOVA and pairwise $t$-tests (Bonferroni correction, $\alpha = 0.0083$) showed PPN significantly outperformed the other methods in all metrics in both conditions ($p < 0.0083$), especially in MSE\textsubscript{amp} and SSIM\textsubscript{amp} ($p < 0.0001$). The three CNN-based methods showed no significant differences, performing similarly across most metrics. Fig. \ref{fig:simu} focuses on intra-family comparisons among deep learning architectures to isolate the impact of structural variations under consistent training supervision. Comparative evaluations with conventional iterative algorithms (e.g., ePIE) are presented separately in Section \ref{sec:traditional_comparison}, where differences in computational efficiency and resilience to varying overlap ratios are quantitatively assessed and visually demonstrated.

\subsubsection{\rev{Spatial Frequency Fidelity Analysis}}  
We analyze frequency-domain fidelity using 1D diagonal cross-sections of the averaged power spectral density (PSD), defining three bands by normalized radial distance: low (0-1/3 Nyquist), mid (1/3-2/3), and high (2/3-1) frequencies.  

Fig.~\ref{fig:psd}(a) reveals PPN closely matches the ground truth's PSD profile, particularly preserving mid-frequency features (red box). CNN-based methods show 2.1$-$3.7$\times$ greater mid-frequency attenuation ($0.06$--$0.07\%$ vs.\ $0.22\%$ total energy) and high-frequency suppression ($<0.03\%$). Quantitatively, PPN achieves $58.6\%$ better mid-frequency preservation than CNNs while maintaining noise-equivalent high-frequency levels ($0.03\%$ vs.\ ground truth's $0.04\%$). This frequency-selective enhancement stems from our polar coordinate attention mechanism, which preferentially weights mid-frequency correlations corresponding to Bragg diffraction conditions while suppressing high-frequency noise through radial decay constraints. The results confirm PPN's unique ability to resolve genuine high-resolution features among the end-to-end methods.

\subsection{Validation on Experimental Synchrotron Data
}

\begin{figure*}[htbp]
\centering
\includegraphics[width=0.9\linewidth]{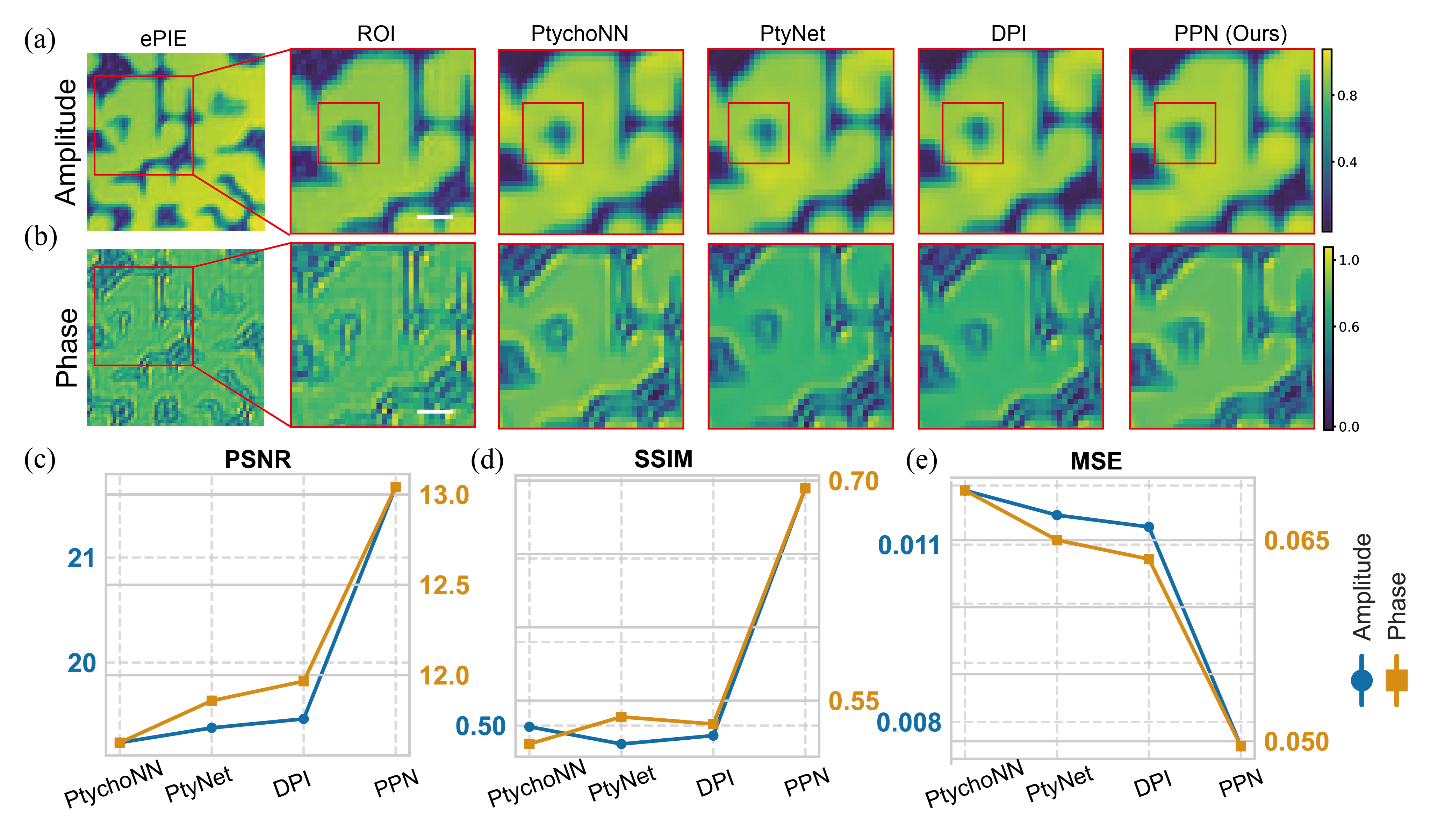}
\caption{Performance comparison on real experimental samples. (a,b) Visual comparison of retrieved amplitude and phase\rev{ (scale bar: 200nm)}. For a specific hook-shaped detail in ROI (region of interest), only our model effectively restores it, significantly outperforming CNN-based methods in fine structure retrieval. (c-e) Quantitative comparison of PSNR, SSIM, and MSE.}
\label{fig:real_samples}
\end{figure*}

\begin{figure*}[htbp]
\centering
\includegraphics[width=0.9\linewidth]{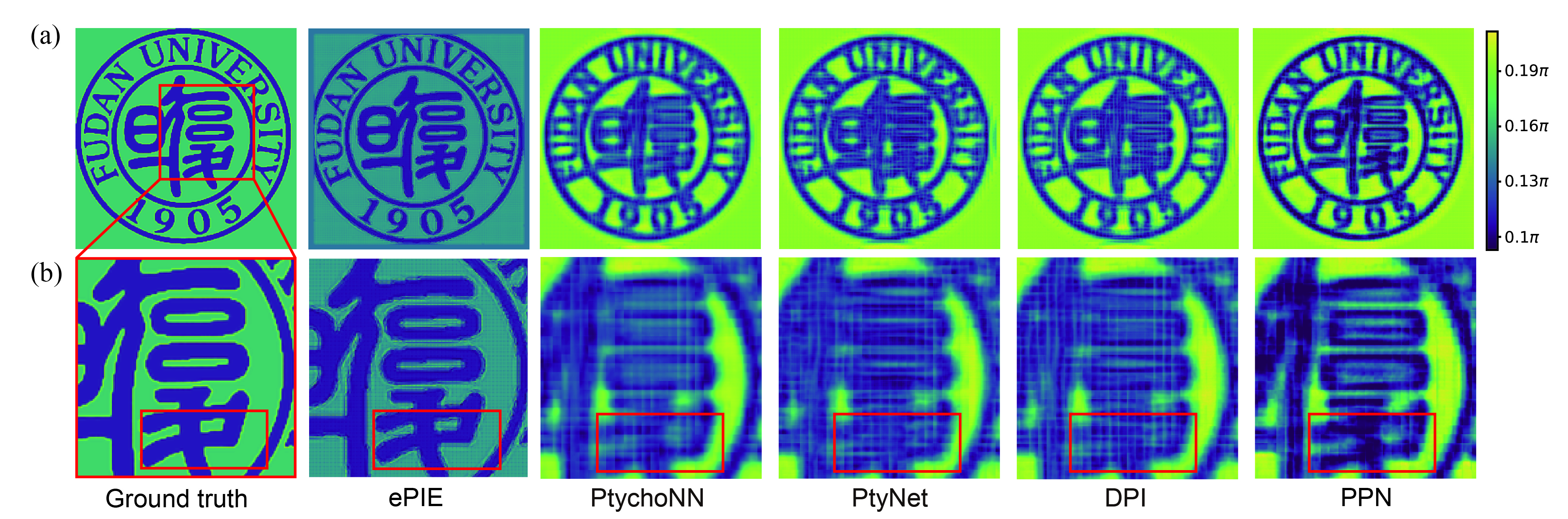}
\caption{Comparison of generalization capabilities across different methods using the Fudan University logo \rev{(simulated data)} as a test sample with distribution significantly different from the training set. (a)~Phase retrieval results with intensity values normalized to $\pi$. (b)~Detailed ROI comparison.}
\label{fig:generalization}
\end{figure*}

Fig. \ref{fig:real_samples} demonstrates PPN's superior performance on real experimental data, consistent with our findings from simulated datasets. Across all metrics (PSNR, SSIM, and MSE) for both amplitude and phase retrieval, PPN consistently outperforms other models. These improvements are visible in fine structural details (Fig. \ref{fig:real_samples} (a,b)). The model's ability to preserve edge sharpness and resolve intricate features in experimental data validates its potential for practical applications where accurate reconstruction of complex nanostructures is essential for scientific interpretation.

\subsection{Generalization Capabilities Comparison}

We evaluated PPN's generalization by training on simulated 1951 USAF pattern (the same with Section~\ref{sec:performance}) with straight-line features and evaluating on complex, curved patterns (Fig.~\ref{fig:generalization}), without fine-tuning, assessing the model's ability to extrapolate learned features to significantly different sample geometries. This evaluation connects directly to practical applications where pre-trained models must process new samples with unknown structures—a scenario where acquiring specific training data is often impractical. In these contexts, generalization capability determines the practical utility of deep learning for ptychographic imaging. Fig.~\ref{fig:generalization} compares phase retrieval results across different methods.\rev{ While deep learning models cannot match the reconstruction quality of the ePIE algorithm, PPN outperforms all other DL-based methods. The red boxes highlight high-frequency details where PPN preserves spatial information with minimal artifacts compared to other end-to-end methods.}

Results show that even with different feature distributions between training and testing data, PPN achieves superior reconstruction quality and better preservation of high-frequency spatial details.

\subsection{Performance Comparison Under Limited Data}
Despite significant advancements in ptychography, a major challenge remains: extended data acquisition periods. While higher overlap between scanning points typically yields better retrieval results \cite{pelzOntheflyScansXray2014}, it also increases experiment duration and radiation exposure. This is particularly problematic for radiation-sensitive materials \cite{bhartiyaXrayPtychographyImaging2021} and in \textit{situ} dynamic studies \cite{wekerSituXraybasedImaging2016}. \rev{Here, the overlap ratio is defined as the physical overlap between adjacent scanning positions in real space.} 

\begin{figure*}[htbp]
\centering
\includegraphics[width=0.83\linewidth]{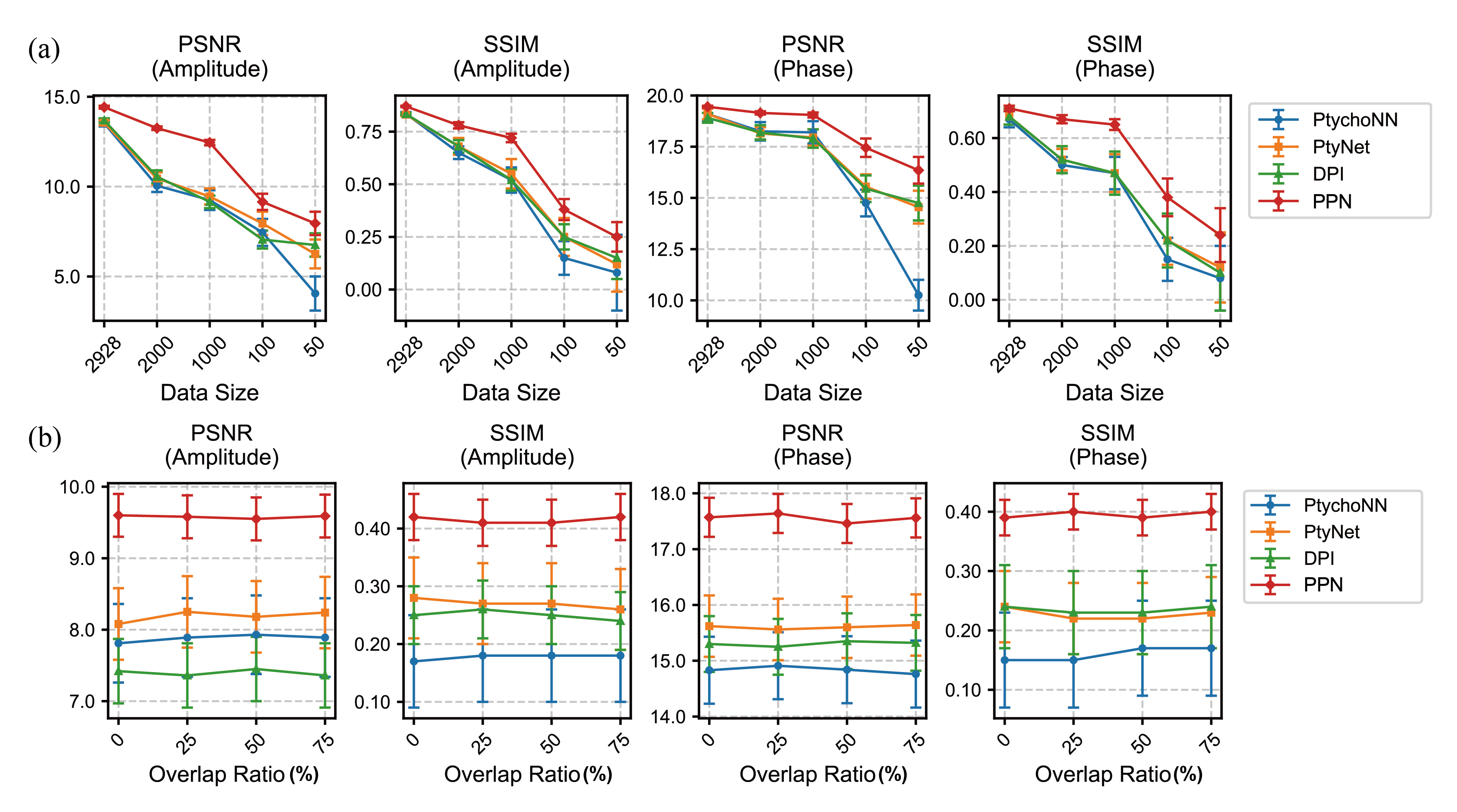}
\caption{Model performance evaluation on simulated data. (a) Reconstruction quality versus training data size at overlap ratio 75\%. (b) Performance stability under different test set overlap ratios with fixed training set size (100 samples).}
\label{fig:size}
\end{figure*}

\subsubsection{Comparison with Different Training Data Size}
\label{sec:datasize}

We evaluated the effect of training data size on model performance by training models with 10 different random initializations while keeping all hyperparameters consistent and testing on the same test set as in Section~\ref{sec:performance}. As shown in Fig.~\ref{fig:size}(a), when the training samples decrease from 2928 to 50, PPN demonstrates superior data efficiency: the PSNR for amplitude reconstruction decreases by only 35\%, and phase reconstruction by 15\%. In contrast, PtychoNN shows the most significant performance degradation, with PSNR dropping by approximately 70\% for both amplitude and phase reconstruction. This resilience, analyzable through compressed sensing theory. \cite{candesRobustUncertaintyPrinciples2006}, suggests PPN learns an optimal sparse prior.

\subsubsection{Comparison Across Overlap Ratios with a Fixed Training Size}
\label{sec:overlap}

We investigated the impact of test set overlap ratios (0\%, 25\%, 50\%, and 75\%) while maintaining \rev{a fixed training set size of 100 samples, consistent with the models used in Section~\ref{sec:datasize}. While the test area maintains consistent spatial coverage (matching Fig.~\ref{fig:simu} (c)'s field of view), the number of test samples decreases with lower overlap ratios due to reduced spatial sampling density.} \rev{As illustrated in Fig.~\ref{fig:size}(b), all models exhibit remarkable metric stability across different overlap ratios. Taking PPN as an example, its amplitude reconstruction PSNR remains stable between 9.5--9.6 ($\Delta=0.1$) and SSIM between 0.39--0.40 ($\Delta=0.01$), with one-way ANOVA ($p > 0.05$) confirming no significant differences between overlap ratio groups.}

\rev{This overlap-invariant behavior stems from our training approach, which uses simulated real-space amplitude and phase as ground truth—data that typically requires exceptionally high overlap ratios to obtain in conventional experimental workflows. This characteristic enables an optimal workflow: using high-overlap data for model training while deploying at low overlap during testing. These findings naturally lead to the question of how PPN compares to traditional iterative methods across different overlap ratios, which we explore in the following section.}

\begin{figure}[htbp]
\centering
\includegraphics[width=0.9\linewidth]{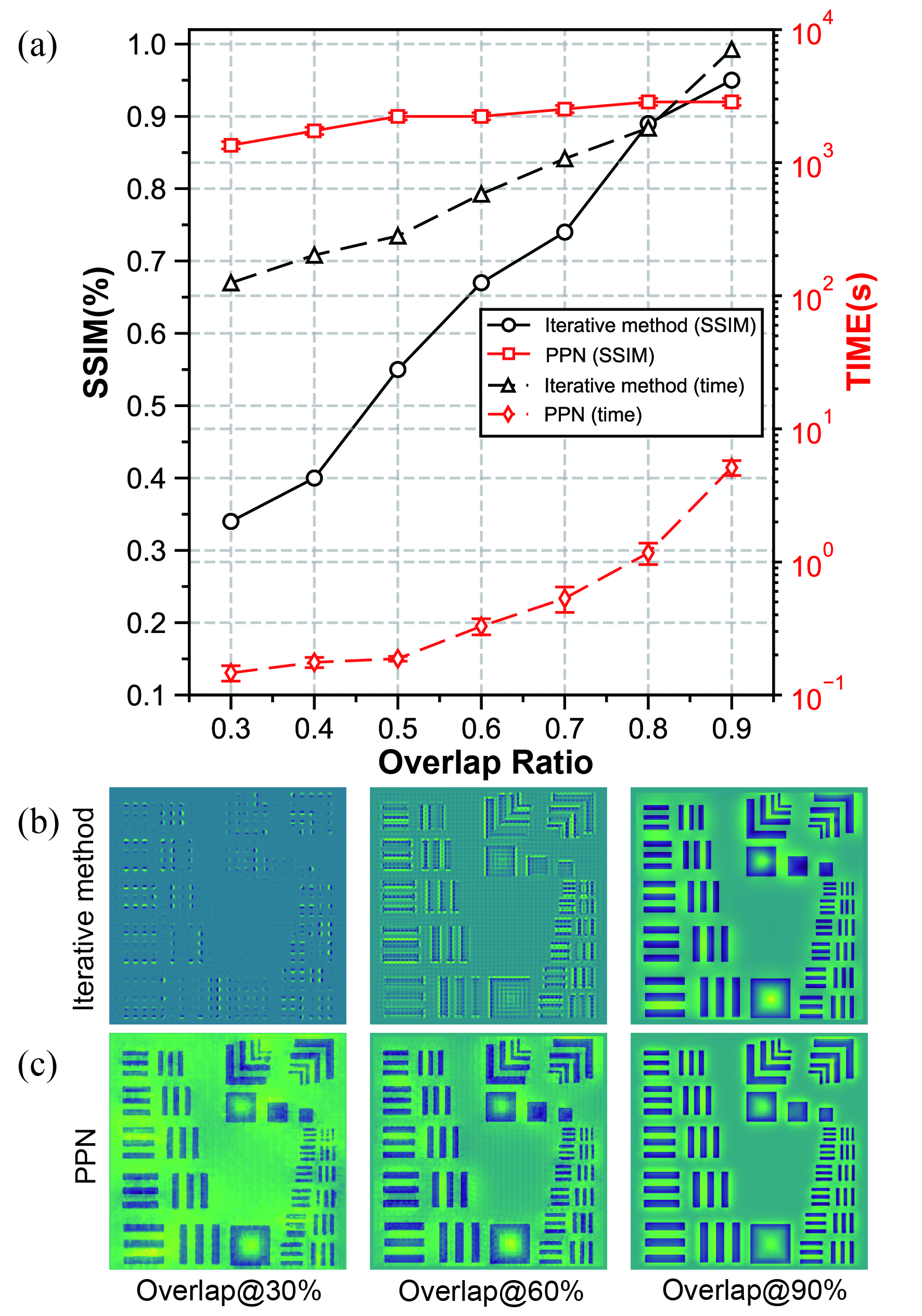}
\caption{Comparison between ietrative method and PPN across different overlap ratios. (a) SSIM and reconstruction time comparison. (b) and (c) Reconstructed images using iterative method and our proposed PPN, respectively, at different overlap ratios (30\%, 60\%, 90\%).}
\label{fig:comparison}
\end{figure}

\begin{figure*}[htbp]
\centering
\includegraphics[width=0.9\linewidth]{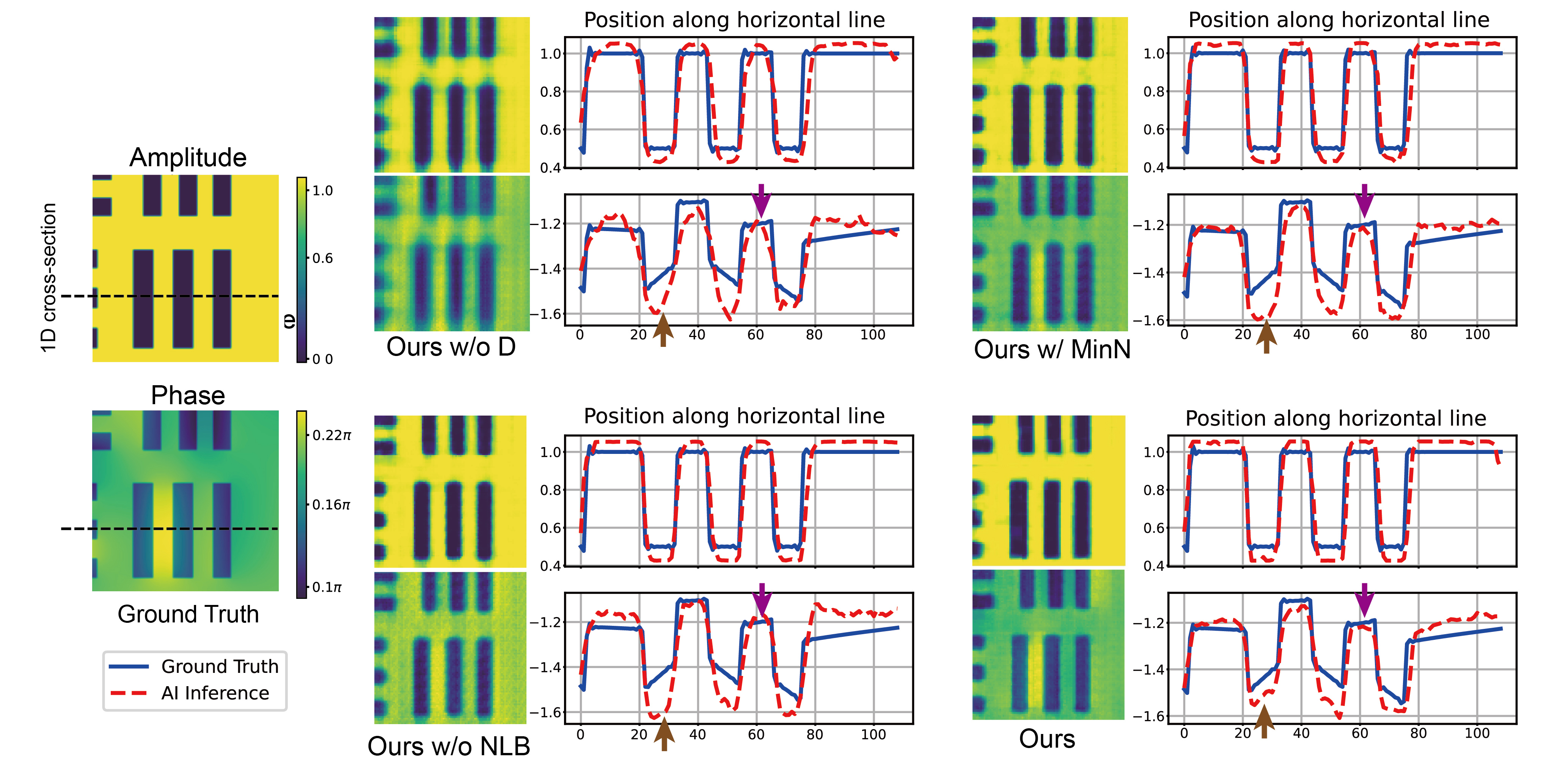}
\caption{Ablation study performed on simulated data showing amplitude and phase reconstructions with 2D images and 1D cross-sectional profiles , comparing ground truth (blue solid lines) with AI inference results (red dashed lines).}
\label{fig:ablation}
\end{figure*}

\subsubsection{Comparison with Iterative Methods Across Overlap Ratios}
\label{sec:traditional_comparison}

\rev{We compared PPN with the iterative method (where we use the widely-used ePIE as our baseline) across various overlap ratios to assess both reconstruction quality and computational efficiency. Fig.~\ref{fig:comparison} presents this comparison, with PPN trained using the same dataset described in Section~\ref{sec:performance}, where sample features remain within the learned distribution. As shown in Fig.~\ref{fig:comparison}(a), PPN maintains SSIM values between 0.86--0.92 across all overlap ratios, indicating consistent reconstruction quality. In contrast, iterative method exhibits strong overlap dependence, with SSIM dropping from 0.95 at 90\% overlap to only 0.12 at 30\% overlap.  At 30\% overlap, PPN completes reconstructions in approximately 0.15 seconds versus iterative method's 125 seconds, representing an 852$\times$ speed improvement, and at 60\% overlap, a 1767× speed improvement. The reconstruction time was measured by averaging over ten consecutive tests, with error bars representing the variance. Reducing overlap from 90\% to 30\% increases data acquisition efficiency by approximately 49 times for a fixed area, as scan point density is inversely proportional to the square of the step size. Theoretically, this combined effect could yield up to 41,748$\times$ (49 $\times$ 852) overall efficiency improvement. This enables significantly faster experiments with substantially reduced radiation exposure while still producing high-quality reconstructions.}

This separation of training and testing strategies, along with PPN's efficiency, demonstrates the synergy between deep learning and domain-specific knowledge in ptychography, providing a viable technical pathway for achieving high-throughput imaging.

\subsection{Ablation Study}

\begin{table}[htbp]
\caption{Ablation Study Results on Simulated Data}
\label{tab:ablation}
\centering
\begin{tabular}{lcccc}
\toprule
\multirow{2}{*}{Model Variant} & \multicolumn{2}{c}{Amplitude} & \multicolumn{2}{c}{Phase} \\
\cmidrule(r){2-3} \cmidrule(l){4-5}
& PSNR (dB) & SSIM (\%) & PSNR (dB) & SSIM (\%) \\
\midrule
Ours w/o NLB & 14.10 & 86.60 & 18.39 & 65.40 \\
Ours w/ MinN & 13.82 & 83.90 & 18.64 & 64.10 \\
Ours w/o D & 11.02 & 60.50 & 17.32 & 54.50 \\
Ours & \textbf{14.42} & \textbf{87.00} & \textbf{19.41} & \textbf{70.60} \\
\bottomrule
\end{tabular}
\end{table}

To justify each component of the proposed PPN, we conducted comprehensive ablation studies. \rev{As shown in Fig. \ref{fig:ablation}, we compared both amplitude and phase reconstructions, including 2D reconstructed images and 1D cross-sectional profiles.} We examined several model variants: 1) Ours without NonLocal Coherence Branch (Ours w/o NLB), which means using only the Local Dependencies Branch and removing the NonLocal Coherence Branch \rev{that contains proposed PoCA}; 2) Ours with Multi-Head Self-Attention (MHSA) in NLB (Ours w/ MinN), replacing \rev{PoCA} with standard MHSA in the NonLocal Branch; 3) Ours without Decoder (Ours w/o D), using fully connected layers instead of CNN for upsampling in the decoder; and 4) Ours (Full Model), the full proposed model.\rev{ The ablation experiments were conducted using the same training set, test set, and parameter settings as in Section~\ref{sec:performance}.}

The quantitative results are presented in Table \ref{tab:ablation}. The removal of the NonLocal Coherence Branch leads to decreased structural coherence in phase reconstruction, which is clearly observable in the 1D profiles. Replacing \rev{PoCA} with standard MHSA results in a 4.16\% drop in PSNR\textsubscript{amp} and 3.56\% in SSIM\textsubscript{amp}, with poorer edge transitions visible in the profile plots. The absence of the CNN decoder causes the most significant performance degradation, with PSNR\textsubscript{amp} and SSIM\textsubscript{amp} decreasing by 23.58\% and 30.46\%, respectively. The reconstructed images without the CNN decoder appear blurry, consistent with the experimental results in \cite{ganPtychoDVVisionTransformerBased2024}. The 1D profiles reveal that without the CNN decoder, the model struggles to maintain accurate amplitude levels and phase transitions. \rev{The full model achieves the best performance across all metrics, particularly in regions with dramatic phase changes in the 1D cross-sectional profiles.} These results validate the effectiveness of each proposed component in our architecture.

\subsection{Model Variants}

To validate our model's effectiveness, we benchmarked against two widely-adopted hybrid CNN-ViT architectures from computer vision: TransUNet \cite{chenTransUNetTransformersMake2021} and SegFormer \cite{xieSegFormerSimpleEfficient2021}. \rev{These baselines remain relevant in recent research across multiple domains \cite{stringerCellpose3OneclickImage2025, chenPIXARTSigmaWeaktoStrong2025, azadMedicalImageSegmentation2024, zhaoFoundationModelJoint2024}, demonstrating their continued utility for important results in various fields.}

The first variant adopts a TransUNet-inspired architecture \cite{chenTransUNetTransformersMake2021}, employing a hybrid CNN-Transformer encoder where CNNs extract initial features before Transformer processing, with a decoder utilizing a cascading structure and skip connections. The second implements a SegFormer-based approach \cite{xieSegFormerSimpleEfficient2021}, featuring a hierarchical structure with progressive resolution reduction for multi-scale feature extraction and replacing the first linear layer in the feed-forward network with a $3\times3$ convolution, \rev{similar to the approach used in PtychoFormer\cite{nakahataPtychoFormerTransformerbasedModel2024}. PtychoFormer uses an encoder based on SegFormer and a decoder adapted specifically for ptychography tasks.} Experiments conducted under noise-free conditions (see Section \ref{subsec:full_stitched_field_retrieval}) demonstrate that PPN outperforms all baseline models (Table \ref{tab:variants}). The PtychoFormer results are comparable to our SegFormer-inspired model due to their similar structures. These variants' limitations stem from designs optimized for real-space image processing: The TransUNet variant, while benefiting from the combination of CNN and Transformer, still relies on local feature extraction in its initial stages. This approach is suboptimal for capturing the global coherence information present in diffraction patterns. Similarly, the SegFormer variant's hierarchical structure, while effective for multi-scale feature extraction in natural images, may lose critical high-frequency information in the context of diffraction patterns due to its progressive downsampling.

\begin{table}[htbp]
\caption{Performance comparison of Model Variants.}
\label{tab:variants}
\centering
\begin{tabular}{lcccc}
\toprule
\multirow{2}{*}{Method} & \multicolumn{2}{c}{Amplitude} & \multicolumn{2}{c}{Phase} \\
\cmidrule(r){2-3} \cmidrule(l){4-5}
& PSNR (dB) & SSIM (\%) & PSNR (dB) & SSIM (\%) \\
\midrule
TransUNet\cite{chenTransUNetTransformersMake2021} & 13.57 & 81.20 & 18.77 & 67.80 \\
SegFormer\cite{xieSegFormerSimpleEfficient2021} & 13.76 & 83.30 & 19.12 & 68.10 \\
\rev{PtychoFormer}\cite{nakahataPtychoFormerTransformerbasedModel2024} & 13.77 & 83.54 & 19.17 & 68.18 \\
PPN (Ours) & \textbf{14.42} & \textbf{87.00} & \textbf{19.41} & \textbf{70.60} \\
\bottomrule
\end{tabular}
\end{table}

\section{Discussion}

\begin{figure}[htbp]
\centering
\includegraphics[width=\linewidth]{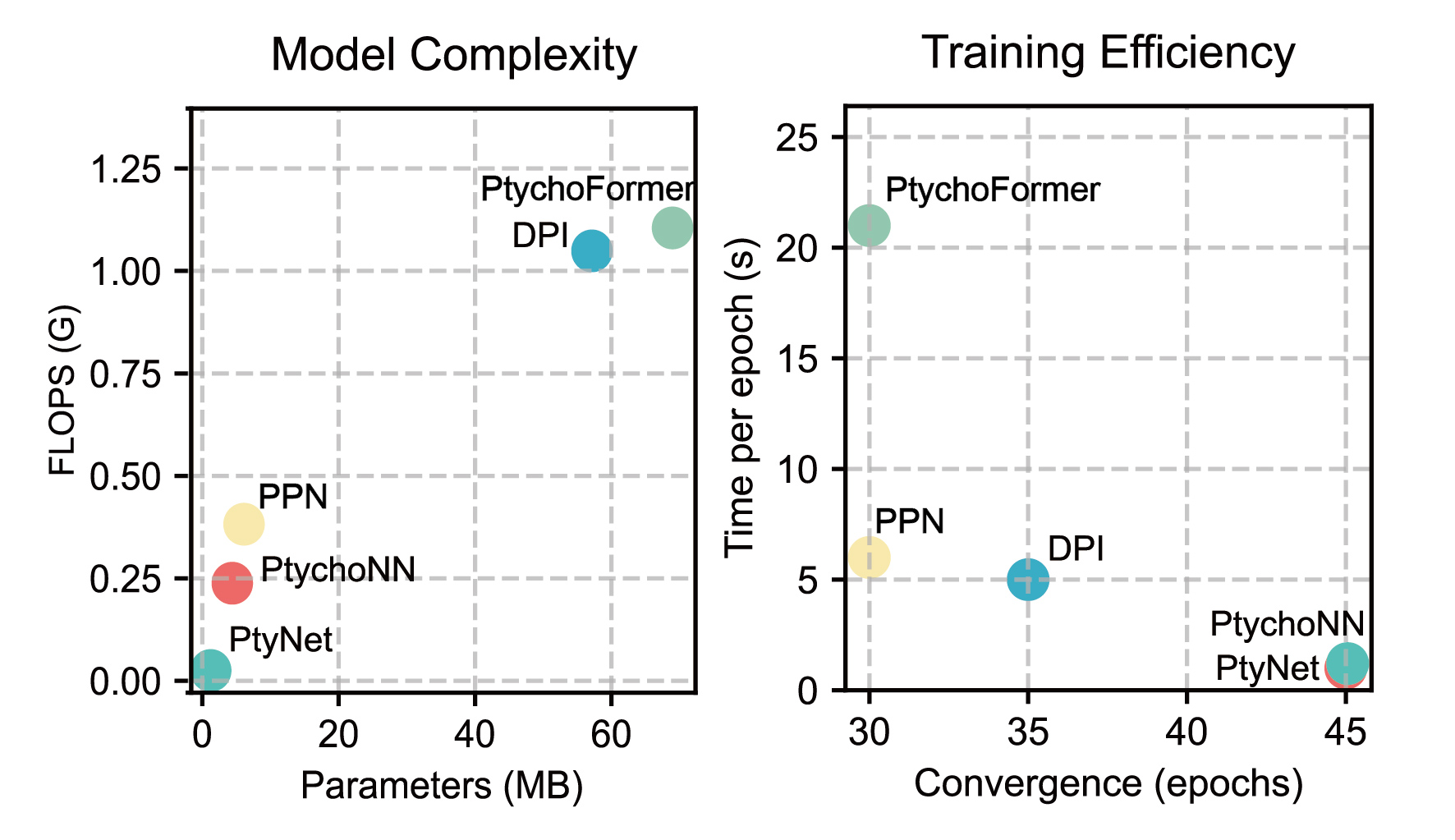}
\caption{Comparison of different models in terms of their complexity (parameters and FLOPS) (left) and training efficiency (convergence epochs and time per epoch) (right).}
\label{fig:cost}
\end{figure}

\rev{\subsection{Computational Complexity Analysis}}
\label{sec:complexity}

\rev{Our analysis compares PPN with CNN-based methods (PtychoNN, PtyNet, DPI) and the ViT-CNN hybrid baseline (PtychoFormer\cite{nakahataPtychoFormerTransformerbasedModel2024}) through computational metrics in Fig.~\ref{fig:cost}. While maintaining hierarchical CNN decoders common to all compared methods, our encoder design achieves substantial efficiency gains.}

\rev{Among CNN approaches, PPN demonstrates intermediate complexity (6.12 MB parameters) between PtyNet (1.24 MB) and DPI (57.14 MB), yet converges faster (30 epochs) than both PtychoNN (45 epochs) and DPI (35 epochs). The 0.382 GFLOPs cost represents a 63.7\% reduction from DPI's 1.049 GFLOPs, confirming efficient feature learning despite transformer integration. Compared to PtychoFormer's ViT-CNN architecture, our model reduces parameters by 91.1\% (6.12 MB vs. 68.96 MB) and FLOPs by 65.4\% (0.382 G vs. 1.105 G) while matching convergence speed (30 epochs). This efficiency comes from three optimizations: 1) fixed feature dimensions preventing channel inflation, 2) parallel dual-branch processing instead of hierarchical refinement, and 3) physics-inspired attention limiting learnable parameters. These changes eliminate redundant operations while preserving physical priors.}

\begin{figure}[htbp]
\centering
\includegraphics[width=\linewidth]{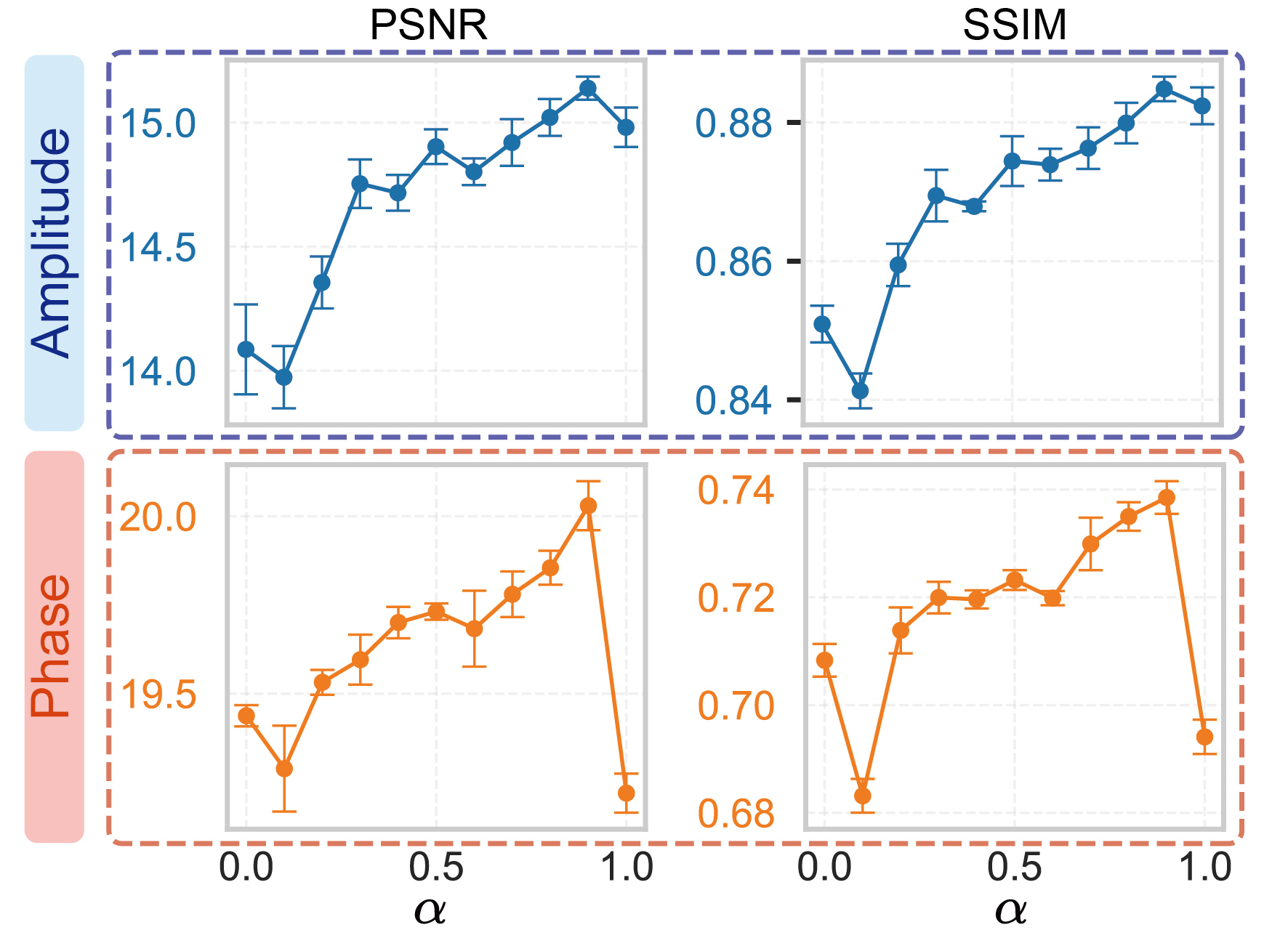}
\caption{Performance analysis of weighted loss function $L_{\text{combined}}(\alpha) = \alpha L_{\text{MSE}} + (1-\alpha) L_{\text{NSSIM}}$ on simulated data. PSNR, SSIM are plotted against $\alpha$ for amplitude and phase retrievals.}
\label{fig:loss_functions}
\end{figure}

\subsection{Loss Function Analysis}

\begin{table}[htbp]
\caption{Comparison with Different Loss Functions}
\label{tab:loss}
\centering
\begin{tabular}{lcccc}
\toprule
Loss Function & \multicolumn{2}{c}{Amplitude} & \multicolumn{2}{c}{Phase} \\
\cmidrule(r){2-3} \cmidrule(r){4-5}
& PSNR (dB) & SSIM (\%) & PSNR (dB) & SSIM (\%) \\
\midrule
MSE Loss & 14.42 & 87.00 & 19.47 & 70.60 \\
MAE Loss & 13.63 & 85.90 & 19.29 & 67.50 \\
Huber Loss & 14.64 & 87.40 & 19.54 & 68.40 \\
NPCC Loss & 10.36 & 69.90 & 17.03 & 66.90 \\
NSSIM Loss & 14.12 & 85.10 & 19.41 & 70.80 \\
\begin{tabular}[c]{@{}l@{}}Weighted Loss \\($\alpha=0.9$)\end{tabular} & \textbf{15.10} & \textbf{88.50} & \textbf{20.04} & \textbf{73.90} \\
\bottomrule
\end{tabular}
\caption*{\footnotesize{Note:The weighted loss combines MSE and NSSIM losses, where $\alpha$ represents the weight of MSE (0.9) and 1-$\alpha$ (0.1) is the weight of NSSIM.}}
\end{table}
To optimize retrieval quality in ptychographic phase retrieval, we evaluated various loss functions (as shown in Table \ref{tab:loss}. Standard metrics like MSE and Mean Absolute Error (MAE) showed similar performance, while the Huber loss  \cite{huberRobustEstimationLocation1992}, defined as: $L_{\text{Huber}}(x) = \begin{cases} \frac{1}{2}x^2 & \text{for } |x| \leq \delta \\ \delta(|x| - \frac{1}{2}\delta) & \text{otherwise} \end{cases}$ offered robustness to outliers. The Negative Pearson Correlation Coefficient (NPCC) \cite{VIINoteRegression1895} loss, despite its success in single-frame phase retrieval \cite{dengLearningSynthesizeRobust2020}, showed poor convergence stability in our experiments. This is likely due to its sensitivity to local statistics in small ptychographic patches, contrasting with its effectiveness on larger, more statistically stable single-frame images.

We introduced a novel combined loss function inspired by the work on image restoration with neural networks \cite{zhaoLossFunctionsImage2016}:  $L_{\text{combined}}(\alpha) = \alpha L_{\text{MSE}} + (1-\alpha) L_{\text{NSSIM}}$ where $L_{\text{NSSIM}} = -\text{SSIM}(I_q, |\mathcal{F}{P(r)\psi(r)}|^2)$, where $\psi(r)$ defined as $\psi(r) = P(r-r_j)O(r)$. This approach balances global consistency (MSE) with local structural preservation ( Negative Structural Similarity Index Measure (NSSIM) \cite{wangImageQualityAssessment2004}). Experiments revealed optimal performance at $\alpha = 0.9$, with consistent improvement as $\alpha$ increased from 0.2 to 0.9, as shown in Fig. \ref{fig:loss_functions}. This robustness to $\alpha$ values reduces the need for precise tuning in practical applications. The combined loss function excels in addressing affine ambiguity, as $\nabla L_{\text{MSE}}$ is sensitive to global affine transformations $T$ ($\frac{\partial L_{\text{MSE}}}{\partial T} \neq 0$), while $\nabla L_{\text{NSSIM}}$ maintains local structure invariance ($\frac{\partial L_{\text{NSSIM}}}{\partial (\text{local structure})} \approx 0$), forming a more robust optimization target.

\section{Conclusion}

We propose PPN, a physics-inspired deep learning framework that significantly improves ptychographic reconstruction through two core innovations: (1) an architecture combining local feature extraction with global pattern coherence modeling, and (2) a novel attention mechanism tailored for diffraction physics. PPN demonstrated superior performance over existing state-of-the-art methods, particularly in high-frequency artifact suppression and data efficiency. Current limitations include performance validation primarily on simulated and small-scale experimental data\rev{ – future work will focus on large-scale real-world deployments across diverse imaging conditions.} The framework's physics-inspired design principles show promising extensibility to other frequency-domain inverse problems like cryo-EM and astronomical imaging.

\section*{Acknowledgments}
We thank the University of Sydney’s Digital Sciences Initiative for financial support of this project through grant DSI Ignite grant scheme.

\bibliography{tci.bib}

\end{document}